\documentstyle{laa}
\def\zabsem{$z_{\rm abs} \approx z_{\rm em}$}
\def\z<<{$z_{\rm abs} \ll z_{\rm em}$}
\def\etal{et al.\,}

\def\lya{Ly$\alpha$\,}

\def\zem{z_{\rm em}}
\def\zabs{z_{\rm abs}}
\def\za{z_{\rm abs}}

\def\kms{{\rm km\,s^{-1}}}
\def\ecs{{\rm ergs\,cm^{-2}\,s^{-1}}}
\def\ang{{\rm \AA}}
\def\apj{ApJ\ }

\def\apjs{ApJS\ }
\def\aa{A\&A\ }

\def\|{$\vert$\thinspace}

\begin{document}
\thesaurus{03(11.01.1; 11.09.3; 11.17.1; 11.17.4 Q0151+048A)}
\title{On the nature of \zabsem\ damped absorbers in quasar spectra
\thanks{Based on observations collected at the
European Southern Observatory, La Silla, Chile.}}
\author{P. M\o ller$^{1,2}$
\and S. J. Warren$^3$
\and J. U. Fynbo$^{1,4}$
}
\offprints{P. M\o ller}
\institute{
$^1$ Space Telescope Science Institute, 3700
San Martin Drive, Baltimore, MD 21218, USA \\
$^2$ on assignment from the Space Science Department of ESA \\
$^3$ Blackett Laboratory, Imperial College of Science Technology and
Medicine, Prince Consort Rd, London SW7 2BZ, UK \\
$^4$ Institute of Physics and Astronomy, University of \AA rhus,
DK-8000 \AA rhus C, Denmark
}
\date{Received 13 May 1997; accepted }
\maketitle
\begin{abstract}

We present spectroscopic observations of the damped \lya absorber at
redshift $z=1.9342$ seen in the spectrum of the quasar Q0151+048A. The
redshift of the absorber is greater than the redshift of the quasar, so
the system resembles the \zabsem\ damped absorber at $z=2.81$ towards
the quasar PKS0528-250. We have previously reported the detection of
\lya emission from the latter absorber, one of only two damped
absorbers for which \lya emission has unambiguously been detected. The
resemblance between the PKS0528-250 and Q0151+048A systems is made
closer by the detection of a weak emission feature in the trough of the
Q0151+048A absorber.
This leads us to
consider whether these \zabsem\ DLA absorbers are different objects to
the intervening DLA absorbers. Two possibilities are examined and
rejected. Firstly the Q0151+048A and PKS0528-250 \zabsem\ absorbers
appear to be unrelated to the intrinsic absorbers (i.e. gas close to
the quasar nucleus, ejected by the quasar), as intrinsic absorbers are
of higher metallicity, have higher ionisation parameter, and show
complex absorption profiles. Secondly these two DLA absorbers cannot be
equated with the gaseous disks of the quasar host galaxies, as the
absorber redshifts differ significantly from the quasar systemic
redshifts. It is likely, then, that intrinsically the \zabsem\ DLA
absorbers are the same as the intervening DLA absorbers, so that
peculiarities in some of the \zabsem\ absorbers can be ascribed to
their different environment i.e. proximity to the quasar, or membership
of the same cluster as the quasar.
We point out that the proximity effect may play some r\^ole, by
reducing the \lya-forest line blanketing of any \lya emission line from
\zabsem\ absorbers.

\keywords{ Galaxies: abundances -- intergalactic
medium -- quasars: absorption lines -- quasars: individual: Q0151+048A}

\end{abstract}

\section{Introduction}

Two optical strategies have been employed to search for high--redshift
normal galaxies. Using narrow--band imaging to detect \lya, relatively
few have been discovered (e.g. Lowenthal \etal 1991,
M\o ller \& Warren 1993 (hereafter Paper I), Francis \etal, 1996).
However, recently Steidel and collaborators
have had considerable success with broad-band imaging, identifying
candidates by the expected Lyman break in their spectra (Steidel \etal,
1996). The same strategy applied to the Hubble Deep Field data has
provided a measurement of the global star formation rate in galaxies in
the redshift interval $2<z<4$ (Madau et al 1997).

The searches for starlight from high--redshift galaxies are
complemented by the analysis of the damped \lya (DLA) absorption lines
in the spectra of high--redshift quasars. The DLA studies have yielded
measurements of the mass density of neutral hydrogen in the universe
(e.g. Wolfe, 1987; Lanzetta \etal, 1991), and the abundance of heavy
elements in the gas (e.g. Pettini \etal, 1994; Lu \etal, 1996), and how
these quantities have changed with redshift. The relation between the
DLA absorbers and the Lyman-break galaxies is not yet well established,
but is important, as it will connect the measured global rate of star
formation with the evolution of the global density of neutral gas, and
its chemical enrichment. This will allow a more detailed comparison
with theories of how galaxies are assembled. For this reason
considerable effort has been devoted to the detection of DLA absorbers
in emission, in order to measure the star formation rate for the
absorber, and the sizes of the cloud of neutral gas, and of the region
of star formation.

Unfortunately to date only two DLA absorbers have been successfully,
and unambiguously, identified. These are i.) the system at $z=2.81$, of
column density ${\rm log(N_{HI})=21.35}$, seen in the spectrum of the
quasar PKS0528-250 (Paper I), and ii.) the system at $z=3.15$, of
column density ${\rm log(N_{HI})=20.00}$, seen in the spectrum of the
quasar Q2233+131 (Djorgovski et al 1996). In the latter case the column
density is below the value usually recognised as defining a DLA system,
but we will treat it as a DLA absorber here. The PKS0528-250 $z=2.81$
DLA absorber has been the subject of extensive imaging and
spectroscopic observations by ourselves. These have yielded clues to the
connection between DLA absorbers and the Lyman break galaxies, described
below. The absorber is nevertheless unusual as the redshift is similar
to the redshift of the quasar. There are only a few such \zabsem\ DLA
absorbers known.\footnote{In this paper we refer to absorbers with
redshifts within a few thousand $\kms$ of the quasar emission redshift
as \zabsem\ absorbers, and to absorbers at lower redshift as intervening
absorbers.} 

The detection of \lya emission from the PKS0528-250 $z=2.81$ DLA
absorber, as well as from two companions with similar redshifts, was
reported in Paper I. The DLA absorber (as well as the companions) has
subsequently been detected in the continuum, with the Hubble Space
Telescope (HST), confirming that the \lya emission is due to star
formation rather than photoionisation by the quasar (M{\o}ller \&
Warren, 1996). The measured half light radius
of the continuum emission, $r_{0.5}=0.13\arcsec$, and the apparent
magnitude, $m_B=25.5$, are within the range measured for Lyman-break
galaxies, which led us to suggest that the two are essentially the same
population (M{\o}ller \& Warren, 1997, hereafter Paper III).
The two companions in this field are similarly small in
size. These HST observations support our earlier suggestion (Warren \&
M{\o}ller, 1996, hereafter Paper II), based on dynamical evidence, that
these three objects are sub-units of a galaxy in the process of
assembly.

Pettini et al (1995) have detected a similar slightly-offset emission
line in the trough of a second \zabsem\ damped absorber at $z=3.083$,
towards the quasar 2059-360. \lya emission may therefore be more
common in or near DLA absorbers near quasars than in or near
intervening DLA absorbers.  Because of the unusual nature of the
PKS0528-250 absorber, and to investigate the possibility that \lya
emission in DLA absorbers is in some way enhanced in \zabsem\ systems,
over intervening systems, we have obtained spectra of a third \zabsem\
DLA absorber, the system at $z=1.93$ seen in the spectrum of the
quasar Q0151+048A. This absorber was first studied by Williams \&
Weymann (1976). The quasar (=UM144=PHL1222, 1950.0 coordinates RA 1 51
17.43, Dec 4 48 15.1) is radio quiet, optically is relatively bright
($m_V=17.63$, $m_B=18.03$), non variable, and has a faint ($m_V=21.2$)
companion quasar Q0151+048B lying 3.3\arcsec to the NE, and at a
similar redshift, discovered by Meylan \etal (1990).

The spectroscopic observations are described in \S2. In \S3 we present
the spectrum. By fitting a Voigt profile to the damped absorption line
we find evidence for an emission line near the base of the trough, just
blueward of the absorption line centre. In \S4 we provide a discussion
of the question of whether the \zabsem\ DLA absorbers are intrinsically
different to the intervening DLA absorbers, and in \S5 we list our
conclusions.

\section{Observations and data reduction}

On the nights of 1994 August 3 and 4 we obtained two spectra of
Q0151+048A, of combined integration time 6200 seconds, using the blue
arm of the EMMI instrument on the ESO 3.5m NTT. We used grism \# 3
(1200 grooves ${\rm mm^{-1}}$, blazed at $3800\ang$) and a $1.5\arcsec$
slit, yielding a resolution of $2.0 \ang$. The detector was a Tektronix
$1024\times1024$ CCD, binned by a factor two in the dispersion
direction, giving a pixel size of $0.38\arcsec\times0.90\ang$. The
readout noise of the chip was $8e^-$.  The journal of observations is
provided in Table 1.

\begin{table}
\caption[ ]{Journal of observations}
\begin{flushleft}
\begin{tabular}{llccr}
\hline\noalign{\smallskip}
Date     & Target     & Exp.  & Seeing & PA      \\
         &            & (sec) &($\arcsec$)&(E of N) \\
3 Aug 94 & Q0151+048A & 2600  & 2.2    &  $90^{\circ}$ \\
4 Aug 94 & Q0151+048A & 3600  & 3.8    & $120^{\circ}$ \\
\noalign{\smallskip}
\hline\noalign{\smallskip}
\end{tabular}
\end{flushleft}
\end{table}

The spectra were reduced using standard techniques for bias and dark
subtraction, and flatfielding. One dimensional spectra were extracted
and combined using the optimising extraction routine described by
M{\o}ller \& Kj{\ae}rgaard (1992). The final spectrum is shown in
Fig. 1.

\begin{figure}
\vspace{8cm}
\includegraphics{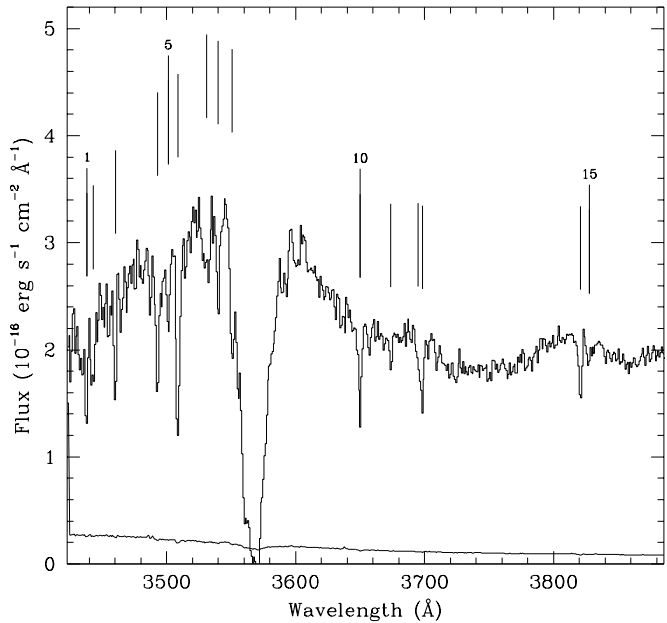}
\caption[ ]{Combined spectrum of Q0151+048A in the region of the DLA
absorption line. Flux per unit wavelength is plotted against
wavelength. The lower solid line shows the $1\sigma$ noise
spectrum. The spectrum has not been corrected for relative slit losses which
were substantial (see text \S3.2).} 
\end{figure}

\section{Results}
\subsection{Absorption lines in the spectrum of Q0151+048A}

The line search algorithm described in M{\o}ller \etal (1994) was used
to find, measure, and identify absorption lines. In this case the
continuum against which absorption lines were measured included the
absorption of a damped Voigt profile near $3567{\rm\AA}$, with
parameters as described below. The resulting line list, excluding the
DLA line, and complete to 4$\sigma$, is provided in Table 2. Listed
there are the observed vacuum wavelength, the observed equivalent width
${\rm W_{obs}}$, and the $1\sigma$ error on the equivalent width
$\sigma_{\rm W}$, for each line. The error on the wavelengths of the
line centroids is dominated by the uncertainty associated with centring
the quasar in the slit, and is estimated to be $0.4\ang$.

\begin{table}
\caption[ ]{Absorption lines in Q0151+048A}
\begin{flushleft}
\begin{tabular}{rcrrcc}\hline
\multicolumn{1}{c}{No.}                 &
\multicolumn{1}{c}{$\lambda_{\rm obs}$} &
\multicolumn{1}{c}{W$_{\rm obs}$}       &
\multicolumn{1}{c}{$\sigma_{\rm W}$}    &
\multicolumn{1}{c}{ID}                  &
\multicolumn{1}{c}{$\za$}                \\
                                        &
\multicolumn{1}{c}{(\AA )}              &
\multicolumn{1}{c}{(\AA )}              &
\multicolumn{1}{c}{(\AA )}               \\ \hline
     1& 3438.4& 0.94&  .18 & \cr
     2& 3443.1& 0.98&  .23 & \cr
     3& 3460.6& 0.97&  .15 & \cr
     4& 3493.0& 1.24&  .14    & SiII (1190)  & 1.9343 \cr
     5& 3501.3& 1.02&  .18    & SiII (1193)  & 1.9342 \cr
     6& 3509.0& 2.24&  .17 & \cr
     7& 3530.9& 1.07&  .17 & \cr
     8& 3539.8& 0.84&  .12 & \cr
     9& 3550.7& 1.13&  .13 & \cr
    10& 3649.8& 1.15&  .13    & SiIV (1393)  & 1.6187 \cr
    11& 3673.9& 0.34&  .09    & SiIV (1402)  & 1.6190 \cr
    12& 3695.1& 0.57&  .11 & \cr
    13& 3698.3& 0.80&  .09    & SiII (1260)  & 1.9342 \cr
    14& 3820.8& 0.76&  .07    & OI   (1302)  & 1.9342 \cr
    15& 3827.4& 0.52&  .12    & SiII (1304)  & 1.9343 \cr \hline
\end{tabular}
\end{flushleft}
\end{table}

Sargent \etal (1988) have identified lines 14 and 15 as the CIV(1548,
1550) doublet at $z=1.468$. With our identification the wavelength
difference between the OI(1302) and SiII(1304) lines, at $z=1.9342$, is
$6.46 \ang$, whereas with their identification the wavelength difference
is only $0.11 \ang$ less. With these data it is not possible to distinguish
between these two possibilities, but given the strength of the other
SiII lines at $z=1.9342$ we find the most natural identification
is that given in Table 2.

The DLA line near $3567{\rm\AA}$ absorbs part of the quasar Ly$\alpha$
and NV emission lines, so to determine the best-fit parameters for the
damped line it was necessary to allow for the fact that the unabsorbed
spectrum (the ``continuum'') is not at all flat in the region of the
DLA line. By dividing the spectrum by the Voigt profile for trial
values of redshift and column density, a first solution for these
parameters was found, which produced a realistic unabsorbed quasar
emission line profile \---\ disregarding the wavelength region over which
the absorption line is saturated. A smooth continuum, including \lya
and NV emission lines, was fitted to this corrected spectrum,
interpolating across the saturated region, $3560\ang-3573\ang$. For
this continuum we now determined the Voigt profile which best fit the
observed spectrum. Dividing again by the model, the procedure was
iterated to a solution. The results of this process are illustrated in
Fig. 2, which shows an expanded plot of the spectrum in the region of
the damped line, together with the best fit to the absorption line.

While there is a certain amount of arbitrariness in the details of the
final model for the quasar emission lines, the same is not true for the
DLA absorption line. The parameters of this line are strongly
constrained by the saturated central part and by the steep sides. The
best fit was obtained with ${\rm N(HI)} = 2.3 \times 10^{20} {\rm
cm^{-2}}$, $\zabs = 1.9342$, but acceptable fits could be obtained for
column densities in the range $2 - 4 \times 10^{20} {\rm cm^{-2}}$.

\begin{figure}
\vspace{8cm}
\includegraphics{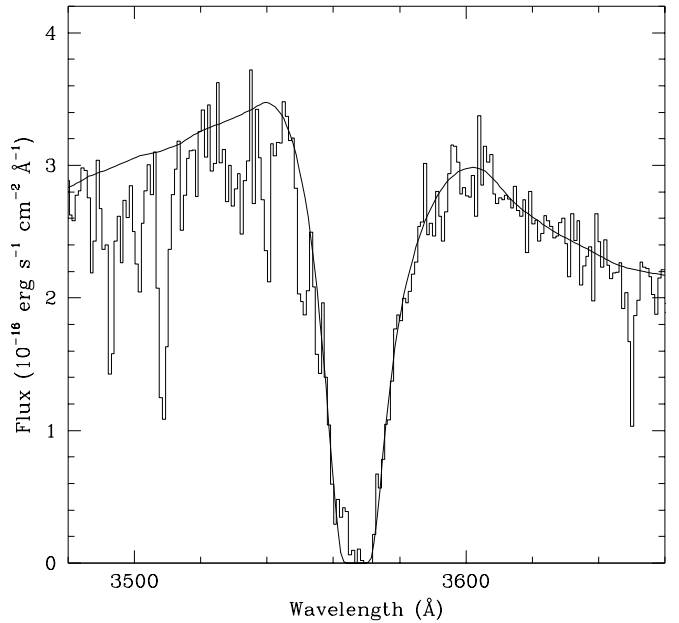}
\caption[ ]{Observed spectrum of Q0151+048A. The smooth line shows
the fit to the damped absorption line, for values of ${\rm N(HI)} = 2.3
\times 10^{20} {\rm cm^{-2}}$, $\zabs = 1.9342$. Note the residual in
the blue side of the saturated part of the DLA line.}
\end{figure}

\subsection{Emission from the DLA absorber}

Inspection of Fig. 2 shows that the model provides a close match to
the profile of the absorption trough, except in the bottom of the DLA
line where there appears to be a weak narrow emission line at the blue
edge of the saturated part of the profile. On the assumption that the
model absorption profile is correct, as evinced by the excellent fit at
all other wavelengths, this emission feature is significant at the
$4.5\sigma$ level. Note that while it is possible to obtain other
acceptable fits to the absorption trough by adjusting slightly the
modelled \lya and NV quasar emission lines, and making corresponding changes
to the DLA parameters, this can never significantly change the
saturated part of the absorption line profile, so the flux in this
emission feature is quite insensitive to the details of the fitting. To
illustrate the emission feature more clearly we have subtracted the
model absorption profile from the data, and divided the difference by
the 1$\sigma$ error spectrum. The resulting residuals (smoothed for
display purposes) are shown in Fig. 3.

\begin{figure}
\vspace{8cm}
\includegraphics{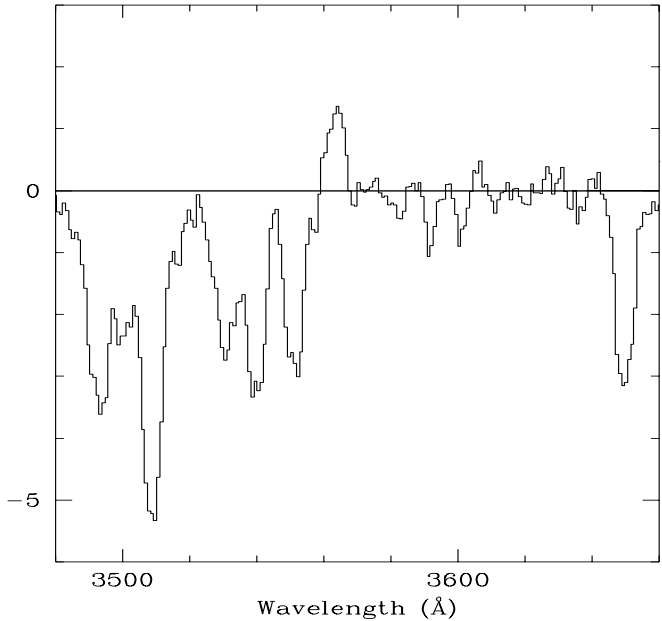}
\caption[ ]{Illustration of the emission feature at 3563.4 $\ang$.
After subtraction of the model absorption profile, the residuals of the
quasar spectrum were divided by the 1$\sigma$ error spectrum. Plotted
is the resultant signal-to-noise spectrum, smoothed with a 5-pixel
boxcar filter. The emission feature, if \lya, is blueshifted relative
to the DLA line by $\approx 300 \kms$.}
\end{figure}

The wavelength centroid of the emission line is 3563.4 $\ang$, and the
measured line flux is $1.2 \pm 0.3 \times 10^{-16} \ecs$. However the
spectrum of Fig. 1 has not been corrected for relative slit losses
between the quasar and the spectrophotometric standard. These were
substantial because the quasar was observed at large zenith
distance. Therefore we calibrated our spectrum by firstly scaling
to the spectrum of Osmer, Porter, and Green (1994), and then
scaling their spectrum to the literature UBV magnitudes.  This brings
the line flux to $\approx 3.5 \times 10^{-16} \ecs$, and indicates
that we only captured about 35\% of the flux.  The new value for the
line flux may still be an underestimate of the total line flux,
dependent on the angular size of the emission-line region relative to
the slit width. If the line is interpreted as \lya emission the
redshift of the object is $\zem =1.9312$, which is $300 \kms$ to the
blue of the absorber. The line luminosity would be several times that
for the DLA absorber towards PKS0528-250 (object S1, Paper I).

\begin{table*}
\caption[ ]{Emission lines and derived emission-line redshifts}
\begin{flushleft}
\begin{tabular}{llccrrc}\hline
Object              &
Ion                 &
$\lambda_{\rm vac}$ &
$z_{\rm em}$        &
W$_{\rm obs}$       &
W$_{\rm rest}$      &
$z_{\rm em}$        \cr
           &      & (\AA )& (peak)     & (\AA )& (\AA ) & TF \\ \hline
Q0151+048A:& \lya$^a$     & 1215.67    &[1.917]& 140    &48& \cr
           & NV$^a$       & 1240.13    & 1.904 &  12.3  &  4.2 & 1.906\cr
           & SiII         &1264.7$\:\:\,$& 1.921 &   6.1  &  2.1 & 1.922\cr
           & OI           & 1304.46    & 1.924 &  11.9  &  4.1 & 1.924\cr
           & SiIV+OIV]$^b$&1399.7$\:\:\,$&[1.898]&  41    &14& \cr
           & CIV$^b$      &1549.1$\:\:\,$& 1.908 &  64    &22& 1.911\cr
           & MgII$^c$     & 2798.74    & 1.921 &        &      & 1.922\cr
\hline
\hline
Q0151+048B:& MgII$^c$     & 2798.74    & 1.936 &        &      & 1.937\cr
\hline
\multicolumn{3}{l}{
$^a${\scriptsize Measured after correction for DLA absorption}} \cr
\multicolumn{3}{l}{
$^b${\scriptsize From Sargent \etal 1988}} \cr
\multicolumn{3}{l}{
$^c${\scriptsize Measured on spectrum in Meylan \etal 1990}} \cr
\end{tabular}
\end{flushleft}
\end{table*}

\subsection{The emission redshifts of Q0151+048A, B}

In considering the nature of the DLA absorber, and the object
responsible for the emission line, it is important to measure
accurately the redshift of the quasar. The Balmer lines or narrow
forbidden lines can be used to measure the systemic redshift, but have
yet to be observed for this object. Instead we must use lines in the
restframe UV. However it is well documented that the high ionization
lines (e.g. CIV here) are blueshifted relative to the quasar systemic
redshift by typically several hundred $\kms$ (e.g. Espey 1989, Tytler
\& Fan 1992, Espey 1997). Any blueshift for the low ionisation lines
(e.g. SiII, OI, MgII here) fortunately is small.

In Table 3 we collect values of the redshift measured for a number of
lines in the spectrum of the quasar Q0151+048A, as well as the MgII
line in the spectrum of the quasar Q0151+048B. All estimates are from
fits to the line peak. The values for the weak SiII and OI lines are
our own measurements from the spectrum of Fig. 2. For both of these
lines we used rest wavelengths of both multiplets under the assumption
that the lines are optically thick. The values for the \lya and NV
lines are also our own measurements from the spectrum of Fig. 2, but
here after division by the model DLA line. This correction does not
strongly add to the uncertainty of the NV emission redshift, but the
error on the \lya emission redshift is dominated by the uncertainty
due to the absorption correction. The data for the
SiIV and CIV lines are taken from Sargent et al (1988). The SiIV
redshift is not very useful however, as the line is chopped up by
strong absorption lines, and it is only included in the table for the
sake of completeness. The values for the MgII lines for the two quasars
were measured by us from the plots of the spectra provided by Meylan et
al (1990).

In a large study Tytler \& Fan (1992) found that, after accounting for
measurement errors, the intrinsic scatter of the blueshift of any
particular line relative to the systemic redshift is small, no more
than $200\kms$, and they tabulated mean values of the blueshift for
several lines. The corrections are smallest for the low ionisation
lines. For example for OI and MgII they found mean values of 50 and 100
$\kms$ respectively. In the last column of Table 3 we list the
redshifts (TF) after applying the corrections suggested by Tytler \& Fan.

As discussed below, the corrections for the high-ionisation lines may
not be suitable for bright quasars. Therefore to estimate the quasar
systemic redshift we have formed a weighted mean of the
redshifts for the three lines SiII, OI, MgII, and added 100 $\kms$ to
the final error as an estimate of the systematic uncertainty. Our best
estimate of the systemic redshift for Q0151+048A is then
$z=1.922\pm0.003$. For Q0151+048B our best estimate for the redshift is
based on the MgII line only, and is $z=1.937\pm0.005$.

From Table 3 it can be seen that for the two high ionization lines,
CIV and NV, the corrections suggested by Tytler \& Fan are too small.
The CIV line in Q0151+048A is blueshifted by $1440\kms$ relative to
the systemic redshift, and this is much larger than the mean blueshift
for this line of $310\kms$ quoted by Tytler \& Fan. Espey (1997), and
M{\o}ller (1997) have found several other cases of quasars where the
CIV line is blueshifted by a similar, or larger, amount. Both Corbin
(1990) and Espey (1997) find a correlation between blueshift of the
CIV line and quasar brightness, and this is likely to be the
explanation for the discrepancy, since the sample of Tytler \& Fan
contains very few optically bright quasars. In fact the correlation is
visible in their Fig. 27. The correlation found by Espey would suggest
a correction of $\approx 1100 \kms$ for the CIV line for Q0151+048A,
bringing it in line with the low ionisation lines.

To summarise, our best estimate for the systemic redshift of Q0151+048A
is $1.922 \pm 0.003$, which is $1250 \pm 300 \kms$ lower than the
redshift of the DLA absorber.

\section{Discussion}

The redshift of the DLA absorber towards Q0151+048A is larger than the
redshift of the quasar, so the system resembles the \zabsem\ damped
absorber at $z=2.81$ towards the quasar PKS0528-250, for which we have
previously reported the detection of \lya emission (\S1). The detection
of emission in the trough of the Q0151+048A absorber therefore makes
the resemblance closer. The line could be \lya emission from the
absorber or from a companion. We defer a detailed discussion of the
nature of the emitter to a subsequent paper, where we will report on
narrow-band imaging observations of the line (Fynbo, M\o ller, \&
Warren, in preparation). However it is interesting to note that Pettini
et al. (1995) have discovered a similar, slightly offset, emission line
in the trough of a third \zabsem\ damped absorber, towards the quasar
2059-360. It appears, therefore, that \lya emission may be more common
in or near DLA absorbers near quasars than in or near intervening DLA
absorbers.  Therefore in this section we firstly consider whether these
\zabsem\ DLA absorbers are representatives of a different population to
the intervening DLA absorbers. Two possibilities are considered; that
the clouds belong to the class of intrinsic absorbers, probably ejected
by the quasar, or that we are seeing the disks of the host galaxies.
Both possibilities are rejected, so it is probable that the
\zabsem\ DLA absorbers are similar to intervening DLA absorbers. This
leads us to consider briefly the likely reason for enhanced \lya
emission near quasars.

In the following we limit ourselves to a discussion of the Q0151+048A
and PKS0528-250 systems, as the relevant information for the quasar
2059-360 has yet to be published.

\subsection{The nature of \zabsem\ DLA systems}

\subsubsection{Intrinsic \zabsem\ systems}
If the \zabsem\ DLA absorbers are different from the intervening
systems, one possibility is that they belong to the class of intrinsic
absorbers, which includes the broad absorption lines (BALs), and the
narrow intrinsic \zabsem\ systems optically thin in the
continuum (Savaglio \etal, 1994;
M{\o}ller \etal, 1994; Hamann 1997), which are possibly related to BAL
systems. Both types of intrinsic absorber typically display complex,
but generally smooth absorption profiles (e.g. Barlow \& Sargent,
1997), whereas both the damped systems under discussion are well fit by
single-component Voigt profiles. Intrinsic \zabsem\ systems also
typically have very high metal abundances, solar or several tens times
solar (Petitjean \etal, 1994; M{\o}ller \etal, 1994; Hamann
1997). The metallicity of the DLA absorber in PKS0528-250, on the other
hand, was measured by Meyer \etal (1989) to be only 12\% solar, and by
Lu \etal (1996) to be 17\% solar. These values are representative of
other DLA absorbers. The metallicity of the Q0151+048A DLA absorber has
yet to be measured.

The intrinsic \zabsem\ systems are also typically characterised by
high ionization parameter. If one were to increase the column density
of such a system to the point where it became optically thick to Lyman
continuum photons, low ionization absorption lines would become
visible. However, the part of the cloud facing the quasar would remain
highly ionized, and one would have a system with mixed ionization
(strong CIV and NV as well as SiII and CII). However neither of the DLA
systems under discussion show strong NV absorption. Therefore, on the basis of
absorption profile, metallicity, and ionisation parameter these two
absorbers appear to be representative of other DLA absorbers, rather than
the intrinsic \zabsem\ systems.

\subsubsection{Quasar host galaxies}

Another possible explanation might be that we are seeing neutral gas in
the quasar host galaxy disk. However for Q0151+048A the quasar systemic
redshift $1.922\pm0.003$ (\S3) and the absorber redshift
$z=1.9342\pm0.0003$ are significantly different. The same is probably
true of PKS0528-250. Here the absorber redshift is $2.8115\pm0.0007$,
which differs from the quasar emission redshift $2.768\pm0.002$,
measured by us from the CIV line, by $3440 \kms$. As discussed in \S3
the systemic redshift of the quasar will be higher than the value
measured from the CIV line. However, if we follow Tytler \& Fan (1992)
the correction is only $310\kms$, whereas Espey's (1997) work would
suggest a correction of no more than $\approx 1500 \kms$. Therefore
these two DLA absorbers do not appear to be the signatures of disks of
the quasar host galaxies.

\subsection{\lya emission and the effect of the quasar}
If, as strongly suggested by the above discussion, the \zabsem\ DLA
systems are the same as intervening DLA systems, the enhanced \lya
emission in or near the \zabsem\ absorbers implies that they occupy
different environments to the intervening systems. The most obvious
explanation that comes to mind is that the emission lines in the
troughs of the Q0151+048A and 2059-360 absorbers, if \lya, are due to
photoionisation by the quasar. However, one can imagine several other
possible explanations for the enhanced emission. For example
gravitational interaction between the quasar and absorber might induce
star formation. In any case the \lya emission from the PKS0528-250
DLA appears to be due to star formation, as we have detected continuum
emission from the absorber, as well as from two \lya emitting
companions (Paper III). This might suggest, instead,
that the explanation for enhanced \lya emission near quasars is that
quasar activity (whatever the cause) is more common in regions where
young galaxies are actively forming stars.

Another factor which could play a r\^ole is the so called proximity
effect (e.g. Bajtlik \etal, 1988). Powerful quasars are able to ionize
the neutral hydrogen in the Lyman forest out to large distances from
the quasar. The effect of this would be to reduce any line blanketing
of \lya emission from galaxies in the vicinity of the quasar. Although
the average line blanketing in the \lya forest of the continuum of a
quasar is only modest at this redshift, $\sim 0.3$, the average line
blanketing of the \lya emission line of a galaxy might be greater as
it would be enhanced by the cloud-galaxy correlation function.

\section{Summary and conclusions}

\begin{enumerate}

\item
In this paper we have presented spectra of the quasar Q0151+048A, and
found evidence for emission in the trough of the \zabsem\ DLA
absorption line at $z=1.9342$.

\item
We have previously reported the detection of \lya emission from the
$z=2.81$ \zabsem\ DLA system towards the quasar PKS$0528-250$, while an
emission line in the trough of a third \zabsem\ DLA absorber, at
$z=3.083$, has also recently been reported. There is only one published
successful detection of \lya emission from an intervening DLA absorber,
so these results suggest that \lya emission is more common in or near
\zabsem\ DLA absorbers than in or near intervening DLA absorbers.

\item
Despite this we find no evidence that \zabsem\ DLA absorbers are not
members of the same population as intervening DLA absorbers. In
particular we are unable to make a connection between the \zabsem\ DLA
absorbers and the so-called intrinsic absorbers, which are of higher
metallicity and higher ionisation. Neither is it possible to associate
the \zabsem\ DLA absorbers with the disk of the quasar host galaxy, as
the redshifts of the absorbers are not compatible with the measured
quasar systemic redshifts.

\item
Star formation is almost certainly the cause of the \lya emission from
one of the three absorbers discussed here. Photoionisation by the
quasar could be the explanation for the other two emission lines, but
this can only be established by more detailed studies, and other
explanations are possible.

\end{enumerate}

\acknowledgements{}
We wish to thank Brian Espey who calculated the emission line redshift
corrections given in sections 3.3 and 4.1.2 using his semi-empirical
correlation, Pat Osmer for supplying us with his spectrum of
Q0151+048A, and Fred Hamann for a number of useful comments on an
earlier version of this manuscript.  JUF gratefully acknowledges
support from the STScI visitors programme.

\end{document}